\documentclass[reprint,prb,twocolumn,showpacs,preprintnumbers,amsmath,amssymb]{revtex4}
\usepackage{graphicx}
\usepackage{dcolumn}
\usepackage{amssymb}
\usepackage{amsmath}
\usepackage{xfrac}
\usepackage{verbatim}
\usepackage{color}

\newcommand{\exciting}{{\usefont{T1}{lmtt}{b}{n}exciting}}

\begin{document}

\title{Dynamic structure factors of Cu, Ag, and Au: A comparative study from first principles}

\author{Audrius Alkauskas}
\author{Simon D. Schneider}
\author{C\'{e}cile H\'{e}bert}
  \affiliation{Institute of Condensed Matter Physics}
  \affiliation{Interdisciplinary Center for Electron Microscopy, Ecole Polytechnique F\'ed\'erale de Lausanne (EPFL), Lausanne, Switzerland}
\author{Stephan Sagmeister}
  \affiliation{Chair of Atomistic Modelling and Design of Materials, Montanuniversit\"{a}t Leoben, Austria}
\author{Claudia Draxl}
  \affiliation{Humboldt-Universit\"{a}t zu Berlin, Institut f\"{u}r Physik, Berlin, Germany}
\date{\today}

\begin{abstract}
We present a comparative theoretical and experimental study of dynamic structure factors (momentum-dependent loss functions)
of the noble metals Cu, Ag, and Au in the energy range of 0$-$60 eV. The emphasis is on theoretical results that are compared with new 
as well as available experimental data. Dynamic structure factors are calculated within the linear-response formalism of
time-dependent density-functional theory,  using the full-potential linearized augmented plane-wave (FP-LAPW) method. 
For the studied energy range, local-field effects are found to be very important for Ag and Au and only marginally relevant for Cu. 
We present an explanation for this surprising behavior. Loss functions of all three metals possess a complex multi-peak structure. 
We classify the features in the loss function as being related to collective excitations, interband transitions, or mixed modes. 
The impact of short-range correlations on the dynamic response functions are 
evaluated by comparing the results of the random-phase approximation to those of the time-dependent local-density approximation. 
Exchange correlation effects are found to be weak for small momentum transfers, but increasingly important for larger momenta. 
The calculated structure factors agree well with experiments.

\end{abstract}
\pacs{
71.10.Ay, 	
71.15.Qe, 	
71.20.Be, 	
71.45.Gm, 	
79.20.Uv 	
}
\maketitle

\section{Introduction}
Electron energy loss spectroscopy (EELS) in the transmission electron microscope (TEM) is a well-established experimental technique
to study the electronic structure of materials. \cite{Egerton_RPP_2009} Energy loss is directly linked to the imaginary part
of the inverse dielectric function, and therefore many other properties of interest can be extracted from these measurements.
Combined with the imaging capabilities of the microscope, as well as optical and X-ray emission spectroscopies that can be performed in 
the TEM, EELS is an excellent tool for materials characterization. In the past decades, the TEM has experienced an impressive 
improvement in its performance. Modern microscopes equipped with field-emission guns and monochromators routinely achieve an 
energy resolution of 0.5 eV or even 0.1 eV, \cite{Egerton_ultra_2007,Rose_STAM_2008} while the resolution in the momentum 
transfer is better than 0.1 \AA$^{-1}$. Equally strikingly, in scanning TEMs with aberration correction units the electron beam 
can attain diameters of sub-\AA ngstrom dimensions. This provides unprecedented opportunities to study electronic structure at the atomic 
scale. \cite{Urban_NM_2009,Muller_NM_2009}

As a first approximation, a spatially-resolved electron energy loss spectrum can be viewed as the weighted convolution of 
momentum-dependent energy loss spectra. Thus, the analysis of spatially-resolved spectroscopic information of complex inhomogeneous 
materials requires the understanding of momentum-resolved electron energy loss spectra of its constituents. Since theoretical 
tools to accurately calculate momentum-dependent dielectric functions have also matured, 
\cite{Hedin,Quong_PRL_1993,Maddocks_EPL_1994,Petersilka_PRL_1996,Fleszar_PRB_1997,Onida_RMP_2002,Sottile_IJQC_2005,Chulkov_CR_2006} 
it is very timely to pose an important question: How do results of state-of-the-art theoretical spectroscopy compare to 
measurements performed in modern TEMs?

In this work, we address this question taking three bulk noble metals, Cu, Ag, and Au, as model systems. We focus on low 
(valence) energy losses. The study of optical and dielectric properties of bulk metals is a mature field in fundamental 
research, and has been such for quite some time. \cite{Pines,Palik} To illustrate this, it suffices to say that the first 
successful theoretical analysis of optical properties of simple metals at the microscopic level, the celebrated Drude model, 
precedes the formulation of quantum mechanics by almost three decades. Transition metals with fully or partially occupied 
$d$ states are certainly more complex than simple metals, \cite{Batson_PRB_1983, Schuelke_JPCM_2001} but their optical properties 
have been studied extensively also for decades, both experimentally and theoretically.
\cite{Ehrenreich_PR_1962,Cooper_PR_1965,Mueller_PR_1967,Morgan_PR_1968,Johnson_PRB_1972,Schluter_ZP_1972,Zacharias_SSC_1976,Otto_SSC_1976,
Stahrenberg_PRB_2001,Marini_PhD,Gurtubay_CMS_2001,Campillo_PRB_1999,Hebert_Ultra_2006,Werner_PRB_2008,Werner_JPCRD_2009,Cazalilla_PRB_2000,
Zhukov_PRB_2001,Marini_PRB_2002,Schone_PRB_2003,Alkauskas_ultra_2010,Glantschnig_NJP_2010}
It may thus seem that no aspect concerning valence excitations is left unknown in these metals. While to a certain extent this 
is certainly true for dielectric properties up to 10 eV (visible, near, mid, and far ultraviolet), a lot less is known about 
excitations with energies of several tenths (up to $~$100) of eV (extreme ultraviolet). Lying between the optical range and shallow 
semi-core and core edges, these excitations are, in some sense, no-man's land. Surprisingly, though being very important in 
the measurements of spatially-resolved electronic response, the momentum dependence of the electron energy-loss spectra in this 
energy range is little understood. Therefore, the main objective of the present work is the study of momentum-dependent EELS 
spectra of Cu, Ag, and Au for energies higher than 10 eV. 


This paper is organized as follows. In Section \ref{theory} the  methodology to calculate response functions
as well as loss functions and dynamic structure factors is outlined. The experimental setup and the post-processing
of raw experimental data is described in Section \ref{expt}. Loss functions for small momentum transfers are 
analyzed and compared to experimental results in Section \ref{loss-q-zero}. 
Momentum-dependent loss functions are presented and analyzed in 
Section \ref{loss-q}. In particular, the dispersion of the low-energy plasmon in Ag is computed and compared with experimental 
data. In Section \ref{conclusions}, the  main results are summarized and the impact of our findings is discussed.

\section{Theory and computational details \label{theory}}
In the case of a periodic solid, the double-differential scattering cross section per unit volume
for momentum transfer $\mathbf{Q}$ and electron energy loss $\omega$ is given by (in atomic units):
\cite{VanHove_PR_1954,Pines}
\begin{equation}
\frac{1}{V}\frac{d^2\sigma}{d\Omega d\omega}=\frac{\gamma^2}{4\pi^2}\frac{k_{\text{f}}}{k_{\text{i}}} \ v^2(\mathbf{Q})
\ s(\mathbf{Q},\omega).
\label{ddscs-eels}
\end{equation}
Here $v(\mathbf{Q})=4\pi/Q^2$ is the Fourier-transform of the Coulomb interaction, $s(\mathbf{Q},\omega)$
is the dynamic structure factor (per unit volume), $\gamma=1/\sqrt{1-v^2/c^2}$ is the relativistic factor for incident electrons,
$k_{\text{i}}$ and $k_{\text{f}}$ are the initial and the final electron momenta ($\mathbf{Q}=\mathbf{k}_{\text{f}}-\mathbf{k}_{\text{i}}$). 
$s(\mathbf{Q},\omega)$ describes quantum-mechanical electron density fluctuations of the physical system \cite{VanHove_PR_1954} and is 
directly related to the macroscopic density response function $\chi_{\text{M}}$:
\begin{equation}
s(\textbf{Q},\omega)=-2 \ \text{Im} \chi_{\text{M}}(\mathbf{Q},\omega).
\label{dsf}
\end{equation}
By definition, $\chi_{\text{M}}$ determines the macroscopic dielectric function via:
\begin{equation}
\varepsilon_{\text{M}}^{-1}(\mathbf{Q},\omega)=1 + v(\mathbf{Q}) \chi_{\text{M}}(\mathbf{Q},\omega),
\label{epsilonM}
\end{equation}
and thus the double differential scattering cross section in Eq.\ (\ref{ddscs-eels}) can be also expressed as:
\begin{equation}
\frac{1}{V}\frac{d^2\sigma}{d\Omega d\omega}= -\frac{\gamma^2}{2\pi^2}\frac{k_{\text{f}}}{k_{\text{i}}} \ v(\mathbf{Q}) \ 
\text{Im} \varepsilon_{\text{M}}^{-1}(\mathbf{Q},\omega).
\label{ddscs-eels2}
\end{equation}
The imaginary part of the inverse dielectric function 
\begin{equation}
L(\mathbf{Q},\omega) = -
\text{Im} \varepsilon_{\text{M}}^{-1}(\mathbf{Q},\omega)
\label{loss-definition}
\end{equation}
is a dimensionless quantity and is traditionally called the loss function. 

The dynamic structure factor is also an important quantity in inelastic X-ray scattering spectroscopy (IXS). In IXS, the 
double-differential scattering cross-section is (in atomic units): \cite{Tischler_pssb_2003}
\begin{equation}
\frac{1}{V}\frac{d^2\sigma}{d\Omega d\omega}= 
\alpha^4\left(\mathbf{e}_{\text{i}}\cdot\mathbf{e}_{\text{f}}\right)^2\frac{\omega_{\text{i}}}{\omega_{\text{f}}}s(\mathbf{Q},\omega),
\label{ddscs-ixs}
\end{equation}
where $\alpha=1/137.035$ is the fine structure constant, $\mathbf{e}_{\text{i}}$ and $\mathbf{e}_{\text{f}}$ 
are polarizations of the incident and the scattered wave. 
In the current paper, we will use the terms dynamic structure factor 
and momentum-dependent loss function interchangeably. They are related through
\begin{equation}
s(\mathbf{Q},\omega) = \frac{2}{v(\mathbf{Q})} L(\mathbf{Q},\omega),
\end{equation}
and the relative merits preferring one over the other depend on the context. For example, the dynamic structure factor is more 
convenient when discussing X-ray scattering, since the double-differential scattering cross section in IXS is directly proportional 
to $s(\mathbf{Q},\omega)$. At variance, the loss function is sometimes more convenient when discussing EELS, even though the 
expression for the double-differential scattering cross section of electrons has an additional prefactor $v(\mathbf{Q})$. 

According to Eq.~(\ref{dsf}), the main quantity that needs to be calculated is the macroscopic 
density-response function $\chi_{\text{M}}$. The macroscopic quantity is related to its microscopic counterpart 
$\chi_{\mathbf{G},\mathbf{G'}}(\mathbf{q},\omega)$ via \cite{Sottile_IJQC_2005}
\begin{equation}
\chi_{\text{M}} (\mathbf{Q},\omega) = \chi_{\mathbf{G},\mathbf{G}}(\mathbf{q},\omega).
\end{equation}
Here $\mathbf{G}$ is the reciprocal lattice vector, and $\mathbf{Q}=\mathbf{q}+\mathbf{G}$, so that $\mathbf{q}$
is confined to the first Brillouin zone. Within the linear-response formulation of time-dependent density functional 
theory (TDDFT), \cite{Petersilka_PRL_1996,Onida_RMP_2002} the microscopic density-response function $\chi$ of the 
interacting many-electron system is related to the density-response function $\chi^0$ of the corresponding non-interacting
Kohn-Sham system through the Dyson equation, which is, symbolically:
\begin{equation}
\chi = \chi^0 + \chi^0 (v + f_{\text{XC}}) \chi.
\label{Dyson}
\end{equation}
$f_{\text{XC}}$ is the exchange-correlation kernel that accounts for all many-body effects, and $\chi^0$ is given as
\begin{eqnarray}
\chi^0_{\mathbf{G},\mathbf{G'}}(\mathbf{q}, \omega) =
\frac{2}{\Omega_0}\sum^{\text{BZ}}_{\mathbf{k}}\sum_{n,n'}\frac{f_{n,\mathbf{k}}-f_{n',\mathbf{k+q}}}
{\epsilon_{n,\mathbf{k}}-\epsilon_{n',\mathbf{k+q}}+\omega+i\eta}
\nonumber \\
\times \langle\psi_{n,\mathbf{k}}|e^{-i(\mathbf{q+G})\mathbf{r}}|\psi_{n',\mathbf{k+q}}\rangle
\langle\psi_{n',\mathbf{k+q}}|e^{i(\mathbf{q+G'})\mathbf{r}}|\psi_{n,\mathbf{k}}\rangle.
\label{chi_0}
\end{eqnarray}
Here, $\psi_{n,\mathbf{k}}$, $E_{n,\mathbf{k}}$, $f_{n,\mathbf{k}}$ are single-particle 
wavefunctions, their eigenvalues, and occupation numbers, respectively;
indices $n$ and $n'$ span all bands, so both so-called resonant and non-resonant terms are 
included in this approach;
$\Omega_0$ is the volume of the unit cell.
While the approach is formally exact, approximations are needed in practice. \cite{Onida_RMP_2002}
In this work, we use two different simple approximations. 
Setting $f_{\text{XC}}$ to 0 yields the random-phase approximation (RPA),
in which only the classical Coulomb field of the induced charge density is 
accounted for when calculating $\chi$. The second approximation is the adiabatic local density approximation
(ALDA or TDLDA), in which $f_{\text{XC}}$
is given by \cite{Onida_RMP_2002}
\begin{equation}
f^{\text{TDLDA}}_{\text{XC}}(\mathbf{r}t,\mathbf{r'}t')=\delta(t-t')\delta(\mathbf{r}_1-\mathbf{r}_2) 
\frac{dV_{\text{XC}}^{\text{LDA}}(n)}{dn}|_{n=n(\mathbf{r,t})}.
\label{ALDA}
\end{equation}
The TDLDA kernel is local in space and time. Thus, its Fourier transform in time
is frequency-independent, and the transform in real space, 
$f^{\text{TFLDA}}_{\mathbf{G,G'}}(\mathbf{q})$, depends only on $\mathbf{q}$ and $\mathbf{G-G'}$.
More complicated kernels, such as those that attempt to include excitonic effects, are not used in this 
work as we are dealing with metals only. 

An important concept in studying response functions is that of crystal local fields. Taking the RPA as an 
example, we obtain from Eq.~(\ref{Dyson}):
\begin{equation}
\chi = \left(1-\chi^0 v\right)^{-1} \chi^0.
\end{equation}
The calculation of $\chi$ therefore involves the inversion of the matrix $1-\chi^0 v$
for each value of $\mathbf{q}$ and $\omega$. Calculations are numerically involved as
one needs to include a sufficient set of reciprocal lattice vectors for the representation of 
microscopic quantities, so that the final result is converged. Neglecting all off-diagonal 
elements, or crystal local fields, is equivalent to dealing with scalar functions rather than matrices.
This is equivalent to the assumption that the screening charge is independent of the position of the 
test charge inside the unit cell. The comparison of response functions calculated
with or without the inclusion of local fields shows to what extent this assumption
is realistic in a given material. 

In this work, the response functions have been determined via all-electron full-potential
calculations based on the linearized augmented plane-wave (FP-LAPW) method, as 
implemented in the \exciting\ code. \cite{Ambrosch-Draxl_CPC_2006,exciting_2005,exciting_2011,Sagmeister_2009} 
For all metals, the experimental lattice constants have been used. Ground-state electron densities,
as well as Kohn-Sham orbitals and eigenvalues have been calculated with the PBE exchange-correlation
functional\cite{Perdew_PRL_1996} (below just referred to as the generalized-gradient approximation, GGA).
The product of the muffin-tin radius and the largest $\mathbf{G}$ vector in the interstitial region,  
$RG_{\text{max}}$, was set to 8.0. An off-center 20$\times$20$\times$20 k-point mesh has been used for 
the Brillouin zone sampling in most cases. For very large momentum transfers we have found it necessary 
to increase the mesh to 25$\times$25$\times$25, while for an accurate determination of the plasmon 
dispersion in Ag a 30$\times$30$\times$30 grid has been employed. To account for local-field effects, 
at least three shells of $\mathbf{G}$ vectors ($\geq$ 27 vectors) have been used. The
summation in Eq.\ (\ref{chi_0}) has been performed with a finite $\eta$, in most cases
0.1 eV.

\section{Experiment \label{expt}}
Thin metallic films have been produced on freshly cleaved NaCl substrates. 
Cu and Ag films have been prepared by sputtering, while the Au film has been produced by evaporation. 
This led to poly-crystalline samples with grain sizes of 10-50 nm and only very little texturing.
The film thickness $d$ was about 55 nm 
for Cu, 50 nm 
for Ag, and 35 nm 
for Au. 
The films have been produced immediately before being transferred to the microscope to avoid oxidation. 
EELS analysis has shown no detectable oxygen contamination.

EELS data for Cu and Au were acquired on a FEI Tecnai F20 TEM equipped with a field-emission gun and a post column Gatan 
imaging filter. The instrumental resolution was about 0.7 eV measured as the FWHM of the zero-loss peak without the specimen. 
The spectra were acquired in the diffraction mode, using a large selected area aperture to integrate over many crystal orientations. 
The momentum transfer was selected by varying the spectrometer's entrance aperture and camera 
length and moving the diffraction pattern projected on the spectrometer entrance aperture
with the projective deflection coils. For Cu, the same camera length and spectrometer entrance aperture was used over the whole 
acquisition range, while for Au the collection angle was increased when moving away from the central spot, allowing for a better 
signal at the expense of a lower angular resolution. The collection semi-angle was 0.1 mrad for Cu and varied from 0.05 to 0.86 mrad 
for Au. The spectra where acquired up to about 2/3 of the distance to the first diffraction ring, since for larger angles the total 
intensity started rising again, an indication that the intensity coming from the first diffraction ring was no longer negligible. 

After the acquisition, the spectra where corrected for multiple scattering following the procedure by Batson and Silcox. 
\cite{Batson_PRB_1983} A spectrum in the image mode needed for this procedure was acquired  using a large collection semi-angle 
of 25 mrad. This spectrum was subtracted from each of the angular resolved spectra after scaling
the intensities of the zero-loss peaks.

Measurements for Ag have been performed with a JEOL 2200 FS TEM,
equipped with a Shottky field-emission gun, an in-column Omega-filter, and a 
2k$\times$2k Gatan Ultrascan CCD camera. The instrumental resolution was 0.8~eV. 
The spectrum was taken in the image mode with a large selection area aperture.

\section{Loss functions at vanishing $\mathbf{q}$  \label{loss-q-zero}}

\subsection{General results}

For small or vanishing momentum transfers and for energies $<$15 eV, loss functions of Cu, Ag, and Au have been 
extensively analyzed before.\cite{Pines,Ehrenreich_PR_1962,Johnson_PRB_1972,Zacharias_SSC_1976,Otto_SSC_1976,Campillo_PRB_1999,
Cazalilla_PRB_2000,Hebert_Ultra_2006,Werner_PRB_2008,Werner_JPCRD_2009,Zhukov_PRB_2001,Gurtubay_CMS_2001,Marini_PRB_2002,
Alkauskas_ultra_2010,Glantschnig_NJP_2010,Marini_PhD} Common features and distinctions for all three metals are rather 
well understood. In particular, the origin of the 3.8 eV plasmon peak in Ag is known, \cite{Pines} and reasons
why a similar excitation does not develop in Cu and is severely damped in Au, have been formulated.
\cite{Cazalilla_PRB_2000} For the sake of consistency we provide a short review of low-energy dielectric properties of 
all three metals in the Supplemental Material.\cite{Supplemental} We discuss the importance of the band structure
in the theoretical description of those properties, complementing the analysis of Cazalilla \emph{et al.} \cite{Cazalilla_PRB_2000} 
from a semi-classical perspective. 
For higher energies ($>$ 10 eV), the features in the loss functions have been 
to some extent addressed for Cu in Ref.\ [\onlinecite{Campillo_PRB_1999}], 
Ag in Ref.~[\onlinecite{Alkauskas_ultra_2010}], and Au in Ref.\ [\onlinecite{Gurtubay_CMS_2001}].

\begin{figure}
\includegraphics[width=7.5cm]{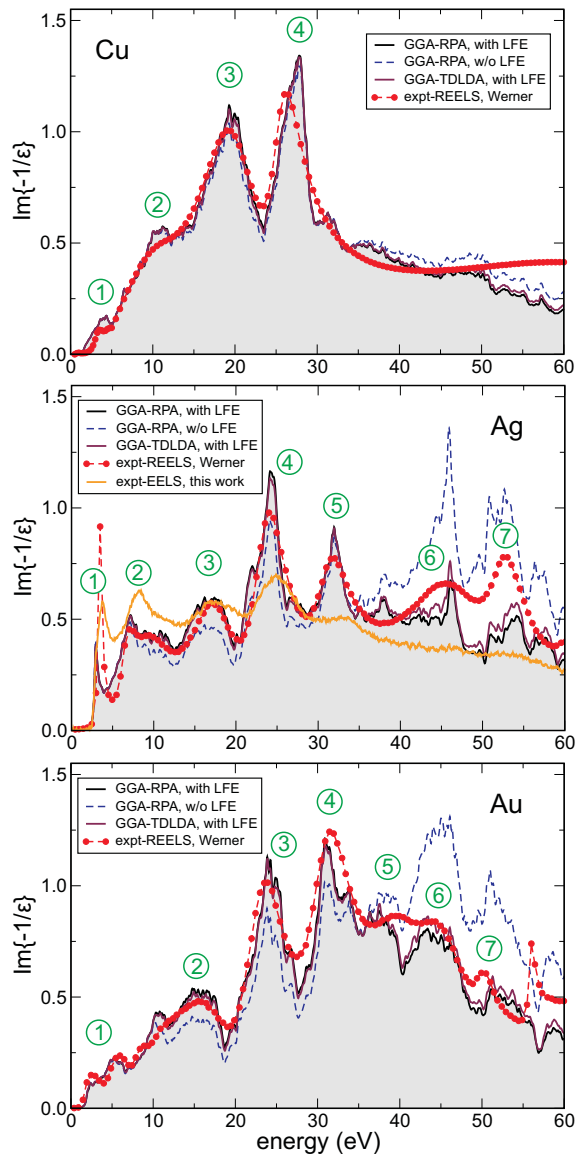}\\
\caption{(Color online)  Loss functions of bulk Cu, Ag, and Au for small momentum transfers based on the GGA band structure 
(energy range 0-60 eV): Black lines and shaded area (blue dashed line) indicate results obtained within the RPA, and include 
(exclude) local-field effects; maroon solid lines are obtained by adiabatic local-density approximation including local-field 
effects. Measurements in the image plane of the TEM are given by orange line (only for Ag); reflection EELS data by Werner \emph{et al.}\cite{Werner_PRB_2008,Werner_JPCRD_2009} are displayed by red dots. 
\label{q-zero-fig}
}
\end{figure}

The calculated loss functions are compared to the experimental data of
Werner \emph{et al.} \cite{Werner_PRB_2008, Werner_JPCRD_2009} in Fig.\ \ref{q-zero-fig}. 
The experimental losss functions were obtained from reflection EELS (disks) using an algorithm 
to separate surface and bulk contributions and to take into account multiple scattering.\cite{Werner_PRB_2008,Werner_JPCRD_2009}
The calculations are all based on the GGA band structure but use three different approximations
to determine response functions: RPA calculations taking local-field effects into account (``GGA-RPA, with LFE'', 
black solid lines and shaded areas), TDLDA calculations with local-field effects (``GGA-TDLDA, with LFE'', solid purple lines), 
and RPA calculations without local field effects (``GGA-RPA, w/o LFE'', dashed lines). Calculations without local field effects 
reproduce earlier results \cite{Werner_PRB_2008,Werner_JPCRD_2009} in which the interband contribution was evaluated strictly for $q=0$, 
and the intra-band contribution had the analytic Drude form with appropriate parameters.

The most important conclusion that can be readily drawn from Fig.\ \ref{q-zero-fig}
is that the two most accurate theoretical treatments (GGA-RPA with LFE and GGA-TLDA with LFE) overall 
provide an excellent desription of loss functions.
More specifically:
(i) In agreement with the established knowledge of low-energy optical properties of these three metals,\cite{Pines,Cazalilla_PRB_2000}
only in Ag does a well-defined low-energy plasmon (peak (1)) develops, while
it is severely suppressed in Cu and Au. This is explained in more detail in Ref.\ \cite{Supplemental}
(ii) Loss functions of all three metals are characterized by a broad structure with several well-defined peaks; 
the origin of the peaks will be discussed in Section \ref{peaks}.
(iii) In the case of Ag and Au, local field  effects are essential for energies $>$ 40 eV.
Only the inclusion of these effects brings the calculated loss functions in agreement with experiment.
At variance, local-field effects are much less pronounced for Cu. 
This surprising asymmetry between metals with very similar nominal electron 
configurations is explained in Section \ref{LFE} and Ref.\ \cite{Supplemental}
(iv)  The effect of including a finite exchange-correlation kernel is approximately an order of magnitude smaller than 
that of the local fields. Consequently, it is almost completely irrelevant for Cu. This will be discussed in more detail in 
Section \ref{LFE}.

\subsection{The origin of peaks in the loss function \label{peaks}}

The origin of the peaks in the loss function below 10 eV has been analyzed before,\cite{Pines,Ehrenreich_PR_1962,Johnson_PRB_1972,Zacharias_SSC_1976,Otto_SSC_1976,Cazalilla_PRB_2000,Alkauskas_ultra_2010} 
and reviewed in Ref.\ \cite{Supplemental}. In short, in Cu peaks (1) and (2) in Cu are characterized by
the a small real part of the dielectric function $\varepsilon_1$, and have to be classified
as plasmon resonances. These resonances are severely damped due to a significant value of the imaginary part of the dielectric
function $\varepsilon_2$. Similar reasoning applies to peak (2) of Ag and peak (1) of Au. \cite{Supplemental}
Ar variance, peak (1) of Ag is a proper plasmon peak that originates from Drude-type oscillations in the $sp$ band renormalized
by interband transitions from the 4$d$ state to the states above the Fermi energy.\cite{Pines,Cazalilla_PRB_2000,Supplemental}
We note in passing that all these peaks have been observed experimentally in numerous occasions. For example, peak (2) in Ag
has been used to image Ag nanoparticles. \cite{Chalker_Nano_2010}

The origin of higher-energy peaks in the loss functions of Cu, Ag, and Au can be understood in terms of 
a system of classical Drude-Lindhard oscillators.\cite{Wilson_PPS_1960} The corresponding model is presented in more detail in Ref.\ 
\cite{Supplemental}.  In short, the analysis shows that for each peak in $\varepsilon_2$
there is an associated peak at slightly larger energies in the loss function, and the difference
between the two frequencies decreases for higher-lying peaks. 
Furthermore, the absolute value of the peak in the loss function is inversely proportional
to the background value of $\varepsilon_1$ at the peak position.

\begin{figure}
\includegraphics[width=8.4cm]{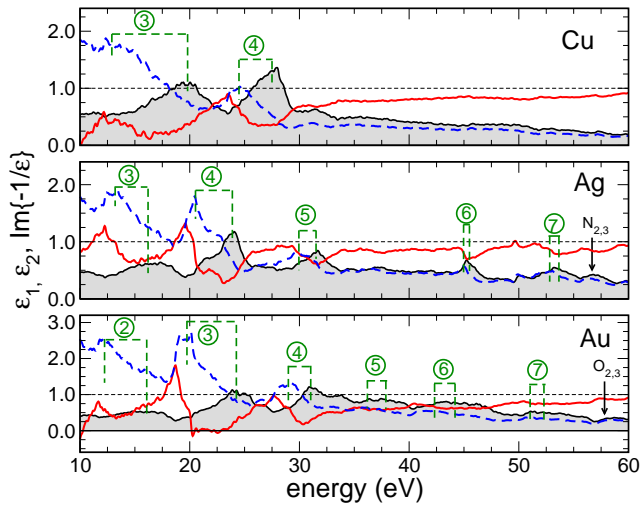}
\caption{$\varepsilon_1$ (solid line), $\varepsilon_2$ (dashed line), and loss function (shaded ares) of Cu, Ag, and Au
in the energy range 11-60 eV. Most important peaks due to excitations of valence electrons are marked in numbers.
N$_{2,3}$ and O$_{2,3}$ edges of Ag and Au are also shown.}
\label{epsilon-large}
\end{figure}

Indeed, as shown in Fig. \ref{epsilon-large}, these trends hold for peaks (3) and (4) of Cu, peaks (3)-(7) of Ag,
and peaks (2)-(7) of Au. This indicates that these excitations originate mainly from interband 
transitions and also implies that the Drude plasmon does not play a significant role at such energies. 

A peak at $\sim$56 eV in the loss function of Ag (Fig.~\ref{epsilon-large}) corresponds to the N$_{2,3}$ core edge 
(excitations from 4$p$ bands), the experimental value being $\sim$62 eV. 
This shows that the 4$p$ states are significantly under-bound in GGA,  even more so than 4$d$ states. Similarly, a peak at 
$\sim58$ eV in the loss function of Au corresponds to the O$_{2,3}$ edge (excitations from 5$p$ bands), while the M$_{2,3}$ 
edge in Cu appears above 60 eV. Since our main focus is on excitations of valence electrons,  however, we will neglect core 
excitations in subsequent discussions.

The conclusion that eminent features in the loss function for energies $>$ 10 eV are caused by interband transitions from $d$ 
states is probably expected and not surprising, and has been already suggested in Ref.\ [\onlinecite{Campillo_PRB_1999}] for Cu. 
The existence of these features on top of a broad background indicates that even at 
energies as high as 50-60 eV above the Fermi level (i.e., more than 45 eV above the vacuum level) electrons still feel the influence 
of the underlying atomic cores. Indeed, for optical transitions from $d$ states to free-electron levels high above the Fermi energy 
one would not expect any sharp features in the loss function. This finding is not new, however. 
In fact, already three decades ago Speier \emph{et al.} performed Bremsstrahlung isochromate spectroscopy (BIS) measurements 
for Cu and Ag.\cite{Speier_PRB_1985} BIS is a variant of the inverse photoemission spectroscopy and measures
the unoccupied density of states. Speier \emph{et al.} found that for energies $>$ 10 eV the unoccupied DOS for Cu and Ag
show several well pronounced peaks. In fact, there is a very nice correlation between peaks in the BIS spectra
and peaks in the EEL spectra. In the case of Ag, for example, for each of the peaks (3)-(7) 
in the EEL spectrum there is a corresponding peak in the BIS spectrum at about 4 eV lower energies.
This nice correspondence underpins the interpretation that pronounced features in the loss
functions arise because of interband transitions from occupied $d$ states to a part in the unoccupied
continuum with a larger density of states.

This conclusions is especially important for photo-electron spectroscopies, 
in which excited photo electrons of energies comparable to these ones are often described by plane waves.

\subsection{Local-field and exchange-correlation effects \label{LFE}}
The loss function (Eq.\ \eqref{epsilonM}) as measured in EELS or IXS, naturally contains LFEs, their strength, however, not  
being experimentally accessible. While the occurence of local-field effects in inhomogeneous solids has been known for decades 
\cite{Adler_PR_1962,Wiser_PR_1963}, their importance varies from solid to solid.
From a theoretical perspective one can get insight by considering the dielectric response of materials with and without these effects. 
where local fields are important both 

It has been demonstrated that the inclusion of LFEs is very important for excitations from core 
\cite{Schwitalla_PRL_1998} and semi-core \cite{Vast_PRL_2002} levels to the 
lowest unoccupied states.  Clearly, localized atomic-like states are far from the homogeneous electron gas model for 
which LFEs vanish. Sturm and Oliveira\cite{Sturm_PRL_1980} have shown, however, that these effects also influence collective 
excitations in simple metals. These excitations couple to short wave-length (large $q$) charge fluctuations via LFEs 
and this significantly affects the width and the dispersion of the resulting plasmon for small $q$. It was subsequently discovered that also the opposite 
effect can be observed: Collective excitations that occur for small $q$ affect electron-hole excitations that occur at very large $q$ 
(beyond the  first Brillouin zone).\cite{Cai_PRL_2006, Errea_PRB_2010}  This interaction is small in Si, where a plasmon Fano 
antiresonance forms,\cite{Sturm_PRL_1992} while in MgB$_2$ \cite{Cai_PRL_2006} and compressed Li \cite{Errea_PRB_2010} the interaction 
is much stronger and leads to replicas of the plasmon at higher Brillouin zones.

Here, we discuss LFE for Cu, Ag, and Au for small $q$. LFE at finite $q$ are analyzed in Section \ref{LFE2}.
The results in Fig.\ \ref{q-zero-fig} show that for small $q$ the inclusion of local-field effects does not lead to notable 
differences for energies $<$ 10 eV, but these effects are very pronounced at larger energies, especially for Ag and Au. 
Since at larger energies, loss functions correspond to transitions from occupied $d$ states to planewave-like unoccupied states
high above the Fermi level, LFEs reflect the rather localized nature of the $d$ orbitals. However, this conclusion seems to contradict
the result that, between 10 and 60 eV, LFEs are substantially smaller in Cu than in Ag and Au. Indeed, Cu 3$d$ 
are more localized than Ag 4$d$ or Au 5$d$ states, and one would naturally expect LFEs effects to reflect this trend.
However, as shown in Ref. \cite{Supplemental}, paradoxically, it is \emph{because} of a larger degree of localization of Cu 3$d$ 
that LFEs are relatively small in the energy range studied. Indeed, we find that LFEs indirectly probe the density of $d$
states in reciprocal space. This density is significantly more spread out for Cu than for Ag and Au. As a result,
it is being probed by planewave-like unoccupied states at higher energies.
This is confirmed by our calculations. However, excitations from semi-core Cu 3$p$  states start to overlap with 
excitations from 3$d$ states at the same energies. The effect of local fields on momentum-dependent 
loss functions is discussed in Section \ref{LFE2}.

It is clear from Eqs.\ \eqref{Dyson} and \eqref{ALDA} that in the optical limit, i.e., $\mathbf{G}=0$ and $q\rightarrow0$, 
the TDLDA kernel does not change the dielectric functions when local field effects are neglected. Indeed, in this case
$\chi_{00}(\mathbf{q,\omega})=\chi^0_{00}(\mathbf{q,\omega})/(1-\left[v(\mathbf{q})+f^{\text{TDLA}}(0)\right]\chi^0_{00}(\mathbf{q,\omega}))$. 
Since $f^{\text{TDLDA}}(\mathbf{G})$ is finite for all $\mathbf{G}$, including $\mathbf{G}=0$, 
and $v(\mathbf{q})=4\pi/q^2\rightarrow\infty$, the effect of the XC kernel is vanishing for small $q$. Thus, the 
choice of the TDLDA kernel is important only when LFEs are included. 
As seen in Fig.\ \ref{q-zero-fig}, the effect of the TDLDA kernel on the loss function is in general small, i.e., much smaller than that of local 
fields. This finding is in line with previous studies. 
\cite{Vast_PRL_2002,Waidmann_PRB_2000} As a result, the effect is totally negligible for Cu for the energies studied here.

\begin{figure}
$ $
\includegraphics[width=8.5cm]{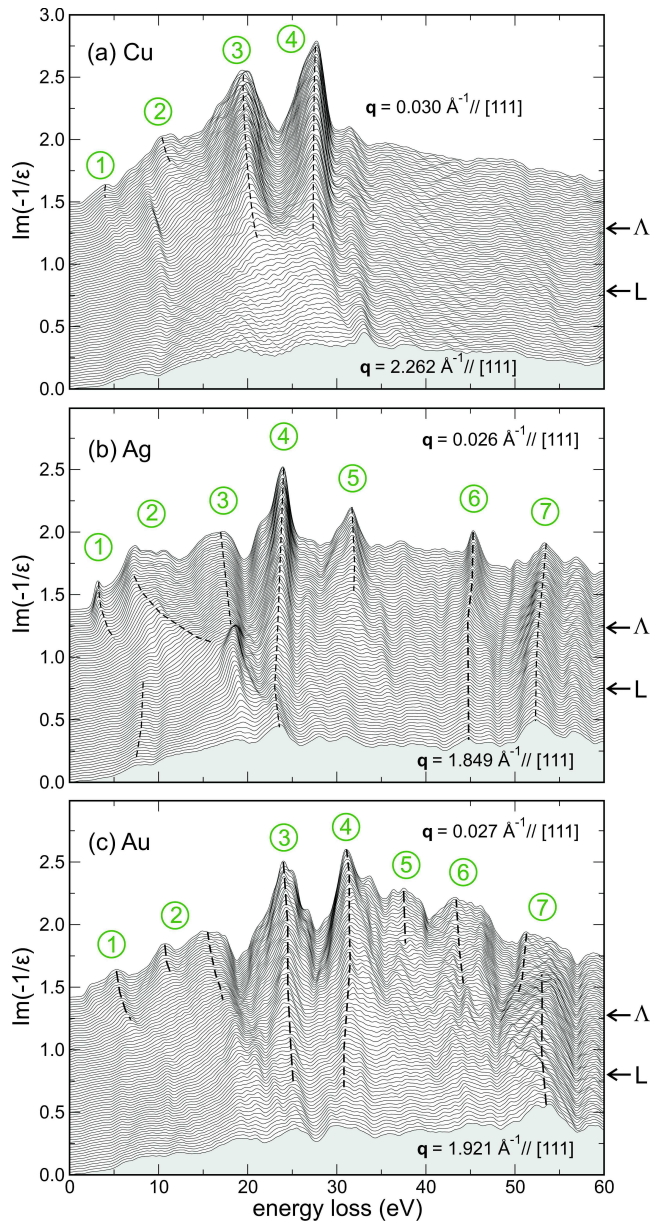}
\caption{Momentum-dependent loss functions of Cu, Ag, and Au in (111) direction.
Spectra are offset for clarity.  Special points $\Lambda=(1/4,1/4,1/4)2\pi/a$ 
and $L=(1/2,1/2,1/2)2\pi/a $ are indicated, where $a$ is the lattice constant.}
\label{loss-q-fig}
\end{figure}

\section{$\mathbf{q}$-dependent loss functions \label{loss-q}}

\subsection{General results}

The calculated momentum-dependent loss functions for all three metals are shown in Fig.\ \ref{loss-q-fig}. 
The momentum transfer is along the [111] direction, and varies 
between 0.030 and 2.262 \AA$^{-1}$ for Cu, 0.026 and 1.849 \AA$^{-1}$ for Ag, and 0.027 and 1.921 \AA$^{-1}$ for Au.
The upper limits correspond to momenta $\mathbf{q}\approx(0.7,0.7,0.7)2\pi/a $, 
where $a$ is the lattice constant. The special points 
$\Lambda=(1/4,1/4,1/4)2\pi/a $ and $L=(1/2,1/2,1/2)2\pi/a $ are indicated in Fig.\ \ref{loss-q-fig}.

Despite several individual features, most of the trends are the same for all three metals.
While the loss functions for small momentum transfers are characterized by several well-pronounced peaks,
as discussed above, their relative intensity tends to decrease with increasing momentum transfer. Qualitatively, 
this can be understood by analyzing the density-response function of non-interacting electrons $\chi^0$ (Eq.\ \eqref{chi_0}).
For small $\mathbf{q}$, $e^{-i\mathbf{qr}}\approx1-i\mathbf{qr}$, and thus only dipole-allowed transitions
contribute, provided local field effects are neglected. Hence, the structure of the loss function is determined mainly by 
transitions from occupied $d$ to unoccupied states with $p$ and $f$ character. For larger $q$, other transitions set in that, 
on average, smear out the peaks. This can also be seen by comparing the loss function for small $q$ 
(mainly dipole-allowed transitions) with the weighted joint density of states for optical transitions, 
defined as $J(E)=1/E\int_{E_{\text{F}}-E}^{E_{\text{F}}}D(E')D(E'+E)dE'$ (not shown). It 
does not discriminate between states of different symmetry. While all the peaks apparent in the 
loss function can be identified in the joint density of states, these are much less pronounced in the latter. 

Another important conclusion that can be drawn from Fig.\ \ref{loss-q-fig} is that peaks caused by interband 
transitions show little dispersion as $q$ increases. Even though for certain peaks the position of their maxima
varies by as much as one electron volt, this is not very significant given that the peak widths range from one to a few eV.

In contrast to peaks that originate from interband transitions, plasmon peaks or peaks that have a plasmon component
(see Section \ref{peaks}) show a different dependence on the momentum. Peak (1) of Ag is the only
well-defined plasmon peak the dispersion of which is clearly parabolic (see Section \ref{plasmon-ag}).
Nevertheless, also other low-energy peaks in Cu, Ag, and Au have a parabolic component in their dispersion. 
Since these peaks are either broad or not well-defined, it is difficult and probably not very useful 
to quantify their dispersion. In the case of Ag, it is interesting to note that one observes a slight increase in 
intensity of peak (2) with momentum transfer where peaks (2) and (3) seem to overlap.

A general trend which can be readily seen in Fig.\ \ref{loss-q-fig} is that, for the energies studied, 
the absolute value of the loss function decreases with increasing momentum. The f-sum rule \cite{Pines} must be fulfilled for 
all $\mathbf{q}$, implying that the weight is redistributed from smaller to larger energies with increasing momentum transfer. 
This aspect is discussed in more detail in Ref.\ \cite{Supplemental}. The reason behind can be already understood by analyzing 
the loss function of the homogeneous electron gas, i.e., the Lindhard dielectric function.

The presented analysis pertains to loss functions that are calculated using GGA single-particle eigenvalues.
The conclusions still hold when approximate $GW$ corrections are applied (figure not shown).
Indeed, we find that only peak (1) of Ag is significantly affected. The intensity and dispersion of peaks that originate 
from interband transitions is barely altered. The only difference is that these peaks move to higher
energies, reflecting the downward shift of $d$ states. This finding underpins that these peaks are caused by optical transitions from $d$ states.

\begin{figure}
\includegraphics[width=7.5cm]{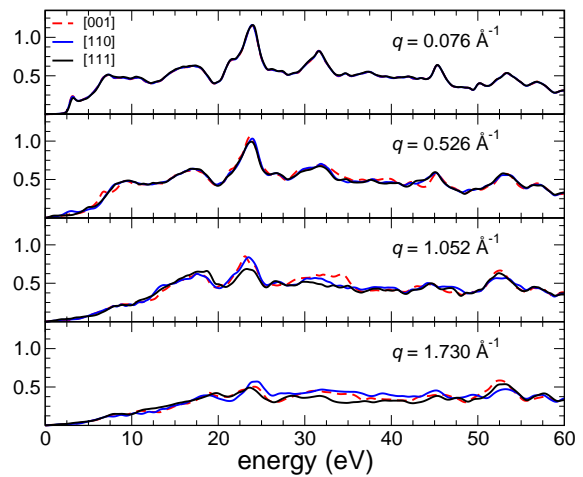}
\caption{Loss functions of Ag for different momentum transfers along high-symmetry crystallographic directions.}
\label{anisotropy-fig}
\end{figure}

\subsection{Anisotropy \label{anisotropy}}

The loss functions presented in Fig.\ \ref{loss-q-fig} correspond to momentum transfer $\mathbf{q}$
in the [111] direction. Due to difficulties in sample preparation, electron microscopy
work is often performed on polycrystalline samples, like those described in Section \ref{expt-res}. 
It is thus important to understand to what extent the conclusions reached in the previous section apply
to other crystallographic directions. 

In Fig.\ \ref{anisotropy-fig}, we show the calculated loss function of Ag for various momentum transfers
(0.076, 0.526, 1.052, and 1.730 \AA$^{-1}$) along the high-symmetry directions [111], [001], and [110].
As expected, the anisotropy is very small for the smallest momenta  and becomes slightly more pronounced
for larger ones. However, even for the largest momentum transfer of 1.730 \AA, differences between 
different crystallographic directions are only quantitative. Indeed, loss functions
for $q$ along different directions possess essentially the same features. 

Despite the lack of very obvious differences, certain minor distinctions can still be identified. 
For example, anisotropy is quite visible for the loss function at about 35 eV, which corresponds to 
peak (5) for smaller momenta (cf. Fig. \ref{loss-q-fig}(b)). The results in Fig.\ \ref{anisotropy-fig}
show that the rate of decrease is slightly different for various crystallographic directions,
being fastest for the [111] direction, slightly slower for [110], and slowest for [001].

We can conclude that the overall anisotropy is quite small for the considered momenta.
Differences between loss functions along different directions are certainly much smaller 
than those between measured and calculated spectra (Section \ref{expt-res}). This, to a 
certain extent, justifies the comparison of loss functions obtained from polycrystalline 
samples with calculations performed for high-symmetry directions.

\begin{figure}
\includegraphics[width=7.5cm]{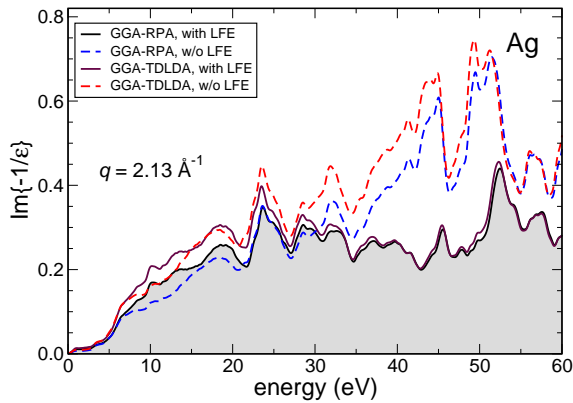}
\caption{(Color online) Calculated Ag loss functions for $q=2.13$ $\text{\AA} $ in the [111] direction.
Black solid line (shaded region): the response function calculated at the RPA level  including
local-field effects (LFE); blue dashed line: the same without LFE; solid purple line:
the response function calculated including the adiabatic LDA kernel (TDLDA) and LFEs; 
red dashed line: the same without LFEs.}
\label{large-q}
\end{figure}

\subsection{Local-field and exchange-correlation effects at finite \boldmath${q}$ \unboldmath\label{LFE2}}

For all three metals that we study here, the  effects of local fields and  the treatment of exchange-correlation effects 
in the response functions for larger momenta are overall quite similar to those at small momenta, as shown in Fig.\ \ref{large-q}. 
Like for small $q$, local-field effects are very pronounced at larger energies, but they seem to kick in already at lower energies.  
The inclusion of the TDLDA kernel has already an effect even without an inclusion of local fields, but in accord with the case $\mathbf{q}\rightarrow 0$, its 
impact is but a fraction of that of the local-field effects. 

\subsection{Comparison with experiment \label{expt-res}}
\begin{figure}
\includegraphics[width=8.5cm]{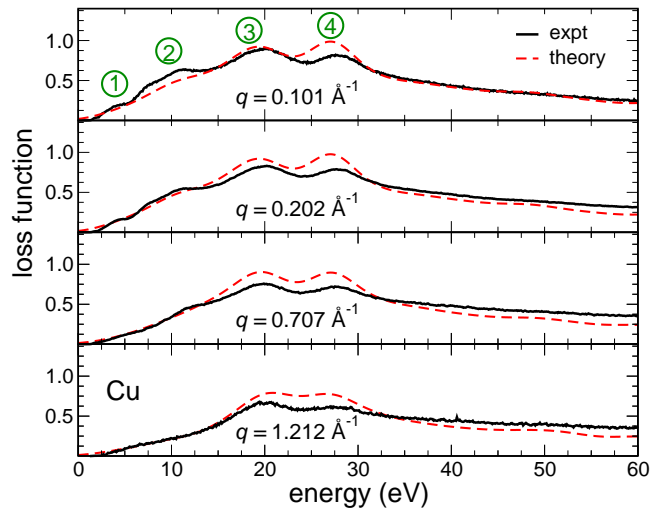}
\caption{Measured (full lines) and calculated (dashed lines) loss functions of Cu for different momentum transfers $q$. 
The experimental spectra are  taken on polycrystalline samples, the calculated data correspond to $q$ in the [111] direction.}
\label{cu-expt-th-1}
\end{figure}

Theoretical and measured momentum-dependent loss functions of Cu for different momentum transfers
are compared in Fig.\ \ref{cu-expt-th-1}. The experimental curves were normalized as described in 
Ref.\ \cite{Supplemental}. Their energy resolution was of the order of one eV, thus an 
additional Gaussian smearing was applied to the theoretical curves for a more meaningful comparison. 
The smearing parameter was kept the same for all momentum transfers and was chosen to obtain the best possible 
overall agreement for all energies and momenta. 
As a matter of fact, the experimental energy resolution was better for smaller momenta. 
This can be explained by longer acquisition times needed for larger momentum transfers.

For the smallest momentum transfer of $q=0.101$ \AA$^{-1}$, the experimental loss functions agrees excellently with 
the calculated one. Peaks (2), (3), and (4), as well as the structure of peak (1) can be identified in the experimental loss function. 
The low-energy peaks are not very apparent in the theoretical curve due to a relatively large smearing parameter.

As explained at the beginning of the present section, the relative intensity of the theoretically obtained peaks decreases with increasing 
momentum transfer for all three metals. This trend is reproduced in the experimental curves. Indeed, for the largest
momentum transfer, $q=1.212$ \AA$^{-1}$,  peaks (3) and (4) are less pronounced than in the case of smaller $q$, 
both in experiment and theory. 

As experimental loss functions tend to acquire more weight at higher energies as compared to their theoretical counterparts, 
the agreement between experiment and theory is slightly getting worse for larger momentum transfers. We assign this fact to an 
incomplete removal of multiple-scattering contributions by the procedure used in the present work. Indeed, the importance of multiple 
scattering increases with increasing scattering angle. However, despite these small drawbacks, we come to the conclusion that 
experiment and theory agree very well with each other. 

\begin{figure}
\includegraphics[width=8.5cm]{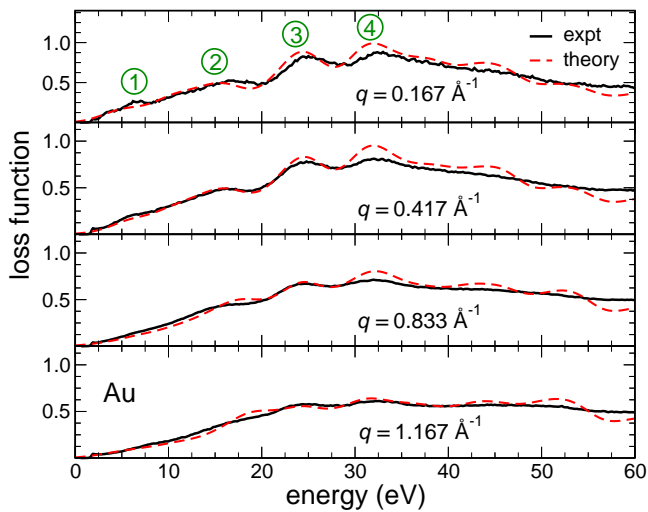}
\caption{Measured (full lines) and calculated (dashed lines) loss functions of Au for different momentum transfers. The experimental spectra
are for polycrystalline samples, the calculated spectra are for $q$ in the [111] direction.}
\label{au-expt-th-1}
\end{figure}

Theoretical and experimental loss functions of Au for various momentum transfers are compared 
in Fig.\ \ref{au-expt-th-1}. While for the smallest momentum transfer, $q=0.167$ \AA$^{-1}$, peaks 
(1), (2), (3), and (4) (cf. Fig.\ \ref{q-zero-fig}) can be identified in the experimental loss functions, 
this is not the case for peaks (5), (6), (7), which are, however, clearly visible in the reflection EELS experiment of 
Ref.\ [\onlinecite{Werner_PRB_2008}] (also Fig.\ \ref{q-zero-fig}). 
Similarly to the case of Cu, the visibility of the peaks decreases with increasing $q$, in full accord with
theoretical results. 

The largest disagreement with theory concerns the visibility of peak (7) for large momentum transfers (Fig.\ \ref{loss-q-fig}(c)). 
Indeed, Fig.\ \ref{loss-q-fig}(c) indicates that peak (7) should be quite pronounced even for large momenta. The results in 
Fig.\ \ref{au-expt-th-1} do not confirm this. One possibility is that multiple scattering contributions
have not been completely removed by the present procedure, like in the case of Cu. However, we
cannot draw a firm conclusion at this moment. Overall the agreement between experiment and 
theory in the case of Au is very good.

Measurements of momentum-dependent loss functions of Ag using a novel technique based on energy-filtered TEM
techniques are in progress in our laboratory. \cite{Schneider_2013}

\subsection{Plasmon dispersion in Ag \label{plasmon-ag}}
In contrast to all other peaks in the loss function, the energy of the plasmon in Ag is very sensitive to 
the accuracy of the band structure. In particular, the position of $d$ states plays a crucial importance.
\cite{Cazalilla_PRB_2000,Marini_PRB_2002,Supplemental} 
One thus would also expect that the same applies to the dispersion of the plasmon. 
Therefore, in calculating properties associated with the low-energy
plasmon we have applied approximate corrections based on $GW$
calculations of Marini \emph{et al.},\cite{Marini_PRB_2002,Supplemental}.
The corrections are functions of energy only. 

\begin{figure}
\includegraphics[width=7.0cm]{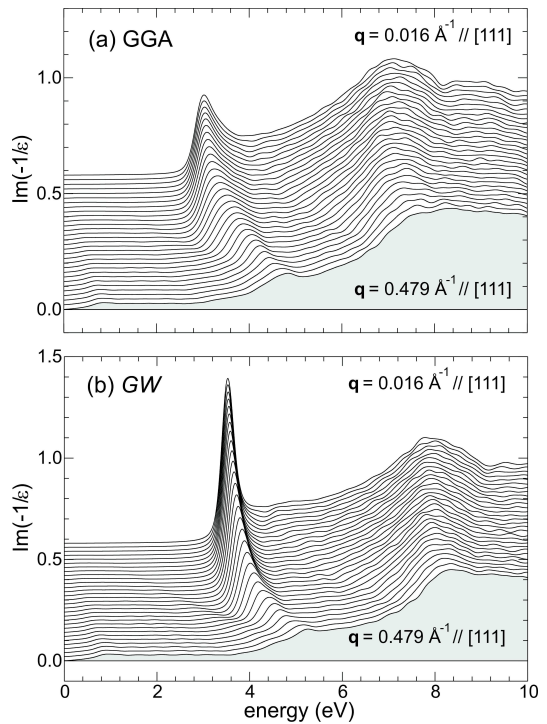}
\caption{Momentum-dependent loss functions of Ag for energies $0-10$ eV calculated within the RPA 
based on the GGA (upper pannel) and the $GW$ (lower pannel	) band structure. Spectra are offset for clarity.}
\label{plasmon_111}
\end{figure}

In Fig.\ \ref{plasmon_111}, we plot the loss functions in the energy range $<$ 10 eV
for momenta $0.016-0.479$ \AA$^{-1}$ along the [111] direction. Spectra 
corresponding to different $q$ are offset for clarity. As $q$ increases,
the plasmon peak moves to higher energies, becomes broader, and gets completely
damped for a critical momentum of $q_\text{c}=0.4-0.5$ \AA$^{-1}$. $q_{\text{c}}$
in Ag is much smaller than in simple metals like Al, where $q_{\text{c}}\approx1.3$
\AA$^{-1}$.\cite{Batson_PRB_1983}
The explanation for this is quite straightforward. In Ag, the plasmon has a much smaller energy due to renormalization
by interband transitions.\cite{Supplemental} Thus, as $q$ increases it
enters the region of electron-hole excitations for much smaller $q$.
This can be expressed by an approximate relationship
$q_{c}\approx\Omega_{\text{p}}/v_{\text{F}}$, \cite{Pines} where $v_{\text{F}}$
is the Fermi velocity.

\begin{figure}
\includegraphics[width=6.5cm]{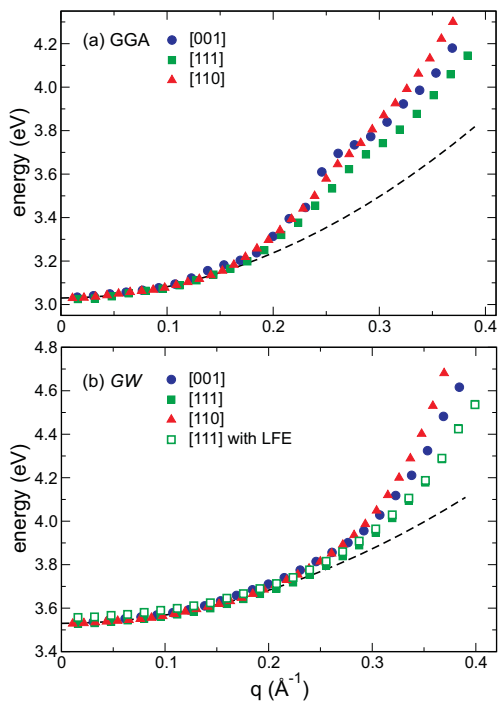}
\caption{Plasmon dispersion in Ag in the three principal crystallographic directions calculated within the RPA 
based on the (a) GGA and the (b) $GW$ band structure. 
}
\label{dispersion_plasmon}
\end{figure}

We have determined the plasmon dispersion by taking the plasmon-peak energy as a function of momentum.
This allows a direct comparison with experimental results, 
where the identical definition is most frequently used.\cite{Zacharias_SSC_1976,Otto_SSC_1976}
The corresponding dependence of the plasmon energy is shown in Fig.\ \ref{dispersion_plasmon}
for three high-symmetry directions. The approximate $GW$ corrections affect not only the position and the width
of the plasmon peak but also its  dispersion. While it is parabolic for low momentum transfers, it departs from this behavior 
for momenta $q>0.2$ \AA$^{-1}$. For the uniform electron gas, it has been shown to be best described by the dispersion relation
$\Omega_{\text{p}}(q) = (\Omega_{\text{p}}^2 + \beta q^2 + \gamma q^4)^{1/2}$.\cite{Jones} However, there is no 
good reason why this function should be used for a more complicated plasmon like that of Ag. Thus, in this work
we choose a different procedure.

In Refs. [\onlinecite{Zacharias_SSC_1976, Otto_SSC_1976}] the plasmon dispersion was measured, and for momenta smaller than 0.4 \AA$^{-1}$ was fitted to a parabola of the following form:
\begin{equation}
\Omega_{\text{p}}(q)=\Omega_\text{p} + \alpha \frac{\hbar q^2}{m}
\label{parabolic}
\end{equation}
Due to  the limited energy resolution and a small number of momentum transfers, the departure
from parabolicity may not have been detected in this experiment. Thus, to compare to these data, we have
also fitted the results of Fig.\ \ref{dispersion_plasmon} to the above given dispersion relation (Eq.~\eqref{parabolic})
in the same momentum range. The results of this fit are presented in Table \ref{plasmon-table}. As mentioned before, 
the plasmon energy at zero momentum $\hbar \Omega_\text{p}$ calculated at the GGA level, departs
significantly from the experimental value, and the agreement is noticeably improved with $GW$ corrections.
Similarly, the dispersion coefficient $\alpha$ is found to be $1.14 \pm 0.10$ in GGA (the error bar here
comes from averaging over the three high-symmetry directions), a bit higher than the measured value of
$0.8 \pm0.1$\cite{Zacharias_SSC_1976} or $0.76 \pm 0.03$.\cite{Otto_SSC_1976} In contrast, the $GW$ value
of $0.85\pm0.10$ agrees very well with the measurements. 

We have also performed a fit including momenta $q<0.2$ \AA$^{-1}$ only
(Table \ref{plasmon-table}), obtaining coefficients $\alpha$ of $0.68 \pm 0.05$ in the case of GGA
and $0.50 \pm 0.05$ for $GW$, respectively. These values could be compared to that of the homogeneous
electron gas having the density of $sp$ electrons in Ag. To do so, requires an additional parameter,
the ``optical'' electron mass, which basically yields the Fermi level relative to the bottom
of the $sp$ band (not to be confused with the effective electron mass $m\approx1$ high
above the Fermi level).
Using $m=0.95m_{\text{e}}$, we obtain $\alpha=0.47$ for the homogeneous
electron gas. This shows that for very small
momenta the plasmon dispersion in Ag does not significantly depart from that in the homogeneous electron gas, 
despite the large difference in the corresponding plasmon energies.

As discussed in Section \ref{loss-q-zero}, local field effects are not very substantial for energies below 10 eV. 
Indeed, when LFEs are included, plasmon energy and dispersion do not change much. This is illustrated in 
Fig.\ \ref{dispersion_plasmon} for calculations including approximate $GW$ corrections.
When local fields are included, the plasmon energy increases slightly from 3.53 eV to 3.56 eV, thus getting
closer to the experimental value of 3.78 eV, while the  plasmon dispersion becomes slightly smaller.
For the [111] direction, the coefficient $\alpha$ decreases by about 3\%. 
To judge about the importance of local field effects, however, more accurate measurements would be desirable.
These should ideally probe also the region where the dispersion departs from the parabolic shape.
We mention here a recent theoretical work by Yan \emph{et al.} for the Ag(111) surface using an orbital-dependent functional. 
\cite{Yan_PRB_2011} Significant improvement was found in the description of the plasmon dispersion in the parabolic region $q<0.15$ \AA$^{-1}$; 
the coefficient $\alpha$, however, was not discussed in that study.

\begin{table}
\caption{Parameters of the plasmon dispersion in Ag for $q<0.4$ \AA$^{-1}$ and for 
$q<0.2$ \AA$^{-1}$.
HEG stands for the homogeneous electron gas with the electron density corresponding to one
electron per Ag unit cell.}
\begin{ruledtabular}
\begin{tabular}{l c c c c}
                                &
\multicolumn{2}{c}{$q<0.4$ \AA$^{-1}$}
&
\multicolumn{2}{c}{$q<0.2$ \AA$^{-1}$}\\
                                &$\hbar \Omega_\text{p}$ &$\alpha$     &$\hbar \Omega_\text{p}$ &$\alpha$       \\
\hline
GGA                             &2.99             &1.14 $\pm$ 0.10&3.03             &0.68 $\pm$ 0.05\\
$GW$                            &3.47             &0.85 $\pm$ 0.10&3.53             &0.50 $\pm$ 0.05\\
Expt. \cite{Zacharias_SSC_1976} &3.78             &0.8 $\pm$ 0.1 &-                &-              \\ 
Expt. \cite{Otto_SSC_1976}      &3.80             &0.76 $\pm$ 0.03&-                &-              \\
HEG                             &-                &-            &-                &0.47            
\end{tabular}
\label{plasmon-table}
\end{ruledtabular}
\end{table}

\section{Discussion and conclusions\label{conclusions}}
In this work, we have performed a comparative analysis of momentum-dependent loss functions (dynamic structure factors)
of the three coinage metals Cu, Ag, and Au. While their dielectric properties have been studied in  
detail for more than half a century, our purpose was to extend these studies to higher energies ($>$ 10 eV)
and larger momentum transfers. As a logical step towards quantitative studies in these less investigated regimes,
we have also discussed the dielectric functions for small energies and have provided additional insight
into the differences and similarities between the three materials.

The main question guiding our work was, to which extent response functions calculated using
state-of-the-art electronic structure techniques, like time-dependent density functional 
theory in the linear-response regime, agree with those measured in modern electron microscopes.
We have shown that in the majority of cases the agreement is indeed excellent. 

Pronounced peaks in loss functions at energies $>10$ eV originate from interband transitions from $d$ states. 
All-electron electronic structure methods like those used in the present work are predestined 
to achieve a reliable description of these features both at small and at finite momentum transfers.

We arrive at the conclusion that the existing theoretical methodologies are indeed able to achieve a very reliable
quantitative description of dielectric properties of the coinage metals Cu, Ag, and Au. It seems that given this excellent 
performance of modern theoretical tools improvements on the experimental side are now necessary to challenge electronic-structure theory. 
This includes better sample quality, post-processing of data, as well as advanced methodologies
to measure momentum-dependent loss functions.\cite{Schneider_2013} Due to strong electron-electron interactions samples for 
TEM measurements should necessarily be very thin. As a result, sample preparation remains one of the most challenging tasks,
in particular regarding means to avoid surface contamination. 
As a step in this direction, new sample preparation and measurements techniques in the case of Ag will be presented 
in a future publication. \cite{Schneider_2013}

\section*{Acknowledgements}
This work has been supported by the Swiss SNF, Project 200021-120308, the Austrian Science Fund, Project P16227, 
and the \"Osterreichische Forschungsf\"orderungsgesellschaft, Project {\it AtoMat}. We acknowledge Y. Ma (Universit\"{a}t 
Osnabr\"{u}ck, Germany) and G.-M. Rignanese (Universit\'{e} Louvain-la-Neuve) for sharing their $GW$ results
of Cu and Au, respectively. We thank C. G. Van de Walle (UC Santa Barbara) for drawing our attention to the relevance of 
some of our results to photoemission experiments, and G. Lucas (EPFL) for carefully reading the manuscript.

\end{document}